\documentclass[11pt,oneside,letterpaper]{article}
\usepackage{amssymb}
\usepackage{amsmath}
\usepackage[dvips]{graphicx}
\usepackage{setspace}
\usepackage{fancyhdr}
\usepackage{ifpdf}
\usepackage{graphicx}
\usepackage{comment}
\usepackage{color}

\DeclareSymbolFont{extraup}{U}{zavm}{m}{n}
\DeclareMathSymbol{\varheart}{\mathalpha}{extraup}{86}
\DeclareMathSymbol{\vardiamond}{\mathalpha}{extraup}{87}

 \def\p{\partial}

\newcommand{\bea}{\begin{eqnarray}}
\newcommand{\eea}{\end{eqnarray}}
\newcommand{\be}{\begin{equation}}
\newcommand{\ee}{\end{equation}}

\newcommand{\bi}{\begin{itemize}}
\newcommand{\ei}{\end{itemize}}

\numberwithin{equation}{section}
\allowdisplaybreaks  

\newcommand{\re}[1]{(\ref{#1})}

\def\eps{\epsilon}

\def\cK{{\cal K}}
\def\cL{{\cal L}}

\def\cQ{{\cal Q}}

\def \eps{{\epsilon}}

\begin{document}

\vspace*{2.5cm}
\begin{center}
{ \LARGE {\textsc{Holography For a De Sitter-Esque Geometry}}\\}
\vspace*{1.7cm}
\begin{center}
Dionysios Anninos$^{\heartsuit}$, Sophie de Buyl$^{\varheart}$ and St\'ephane Detournay$^{\varheart}$
\end{center}
\end{center}
\begin{center}
$^{\heartsuit }$\textit{Center for the Fundamental Laws of Nature, Harvard University, \\
Cambridge, MA 02138, USA }\\
$\;$\\
$^{\varheart}$\textit{Service de Physique th\'eorique et math\'ematique, \\ Universit\'{e} Libre de Bruxelles, Campus Plaine, \\
CP231, B-1050 Brussels, Belgium}
\vspace*{0.8cm}
\end{center}
\vspace*{1.5cm}
\begin{abstract}
Warped dS$_3$ arises as a solution to topologically massive gravity (TMG) with positive cosmological constant $+1/\ell^2$ and Chern-Simons coefficient $1/\mu$ in the region $\mu^2 \ell^2 < 27$. It is given by a real line fibration over two-dimensional de Sitter space and is equivalent to the rotating Nariai geometry at fixed polar angle. We study the thermodynamic and asymptotic structure of a family of geometries with warped dS$_3$ asymptotics. Interestingly, these solutions have both a cosmological horizon and an internal one, and their entropy is unbounded from above unlike black holes in regular de Sitter space. The asymptotic symmetry group resides at future infinity and is given by a semi-direct product of a Virasoro algebra and a current algebra. The right moving central charge vanishes when $\mu^2 \ell^2 = 27/5$. We discuss the possible holographic interpretation of these de Sitter-esque spacetimes.

\end{abstract}

\newpage
\setcounter{page}{1}
\pagenumbering{arabic}
\onehalfspacing

\section{Introduction}

One of the interesting open questions concerning the holographic principle has been the search for examples in spacetimes other than anti-de Sitter space. A prime candidate is de Sitter space, the maximally symmetric cosmology whose natural boundary resides at future and past infinity $\mathcal{I}^\pm$. Given the difficulty in constructing de Sitter vacua in string theory, which are always non-supersymmetric, one must also rely on additional theoretical techniques to analyze them.

A particularly effective toolkit has been that of lower dimensional gravity, where it was found that the asymptotic symmetry group of three-dimensional de Sitter space consists of two copies of the infinite dimensional Virasoro algebra \cite{Brown:1986nw,Strominger:2001pn}. This gave circumstantial evidence for the proposal that quantum gravity in a three-dimensional de Sitter background is holographically dual to a two-dimensional conformal field theory living at $\mathcal{I}^\pm$ (see \cite{Anninos:2010zf} for a discussion in four-dimensions).

Furthermore, it is interesting in its own right to examine the possibility of self-consistent theories of pure gravity\footnote{By pure theories of gravity we mean theories consisting solely of metric degrees of freedom.} at the quantum mechanical level \cite{Witten:2007kt}. So far, there is essentially one non-trivial theory of pure gravity with a viable candidate. The theory is known as topologically massive gravity (TMG) \cite{Deser:1981wh,Deser:1982vy} and is given by the regular Einstein-Hilbert action with positive Newton constant $G$ and negative cosmological constant $\Lambda = -1/\ell^2$, supplemented by a gravitational Chern-Simons term with coefficient $1/\mu$. The candidate is AdS$_3$ at the chiral point $\mu\ell=1$ \cite{Li:2008dq,Maloney:2009ck}, where one observes that under certain consistent boundary conditions, the left-moving sector of the theory classically decouples resulting in a holomorphic theory. It is conjectured that no pathological degrees of freedom arise at the quantum level.


Away from the chiral point $\mu \ell = 1$, where AdS$_3$ is known to be unstable, TMG possesses other candidate stable vacua -- the warped AdS$_3$ vacua \cite{Anninos:2008fx}. Warped AdS$_3$ (with $\mu^2 \ell^2 > 9$) has been conjectured to be dual to a two-dimensional CFT with unequal central charges $c_R$ and $c_L$ \cite{Anninos:2008fx}. The original arguments leading to these conjectures are based on the thermodynamic properties of warped AdS$_3$ black holes, in particular matching the geometric entropy with the one obtained by applying the Cardy formula. Further evidence includes independent computations of $c_R$ and $c_L$, determination of consistent boundary conditions, and computation of boundary correlators and quasi-normal modes \cite{Blagojevic:2009ek,Compere:2008cv, Compere:2009zj,Anninos:2009zi,Chen:2009hg}. In \cite{Lashkari:2010iy} decays from AdS$_3$ to warped AdS$_3$ were studied using the Ricci-Cotton flow.


In this work we will consider topologically massive gravity with a positive cosmological constant $\Lambda = +1/\ell^2$. This theory is known to have various solutions \cite{Nutku:1993eb,Gurses,Chow:2009km,Ertl:2010dh}. We will study a geometry given by a real line fibration over dS$_2$ with an $SL(2,\mathbb{R})\times U(1)$ isometry group appearing in the parameter region $\mu^2\ell^2 < 27$ \cite{Anninos:2009jt}. This geometry has various characteristic features of a de Sitter universe. Observers are surrounded by a cosmological horizon and observe Hawking radiation, and the natural boundary of the spacetime resides again at $\mathcal{I}^\pm$. Unlike dS$_3$ however, global identifications of this geometry, which we refer to as warped dS$_3$, give a two-parameter family of solutions containing two horizons with different Hawking temperatures which are free of closed timelike curves and conical singularities. Furthermore, they asymptote to global warped dS$_3$.\footnote{Warped dS$_3$ is the de Sitter analogue of warped AdS$_3$ \cite{Anninos:2008fx}. It further appears as the near horizon geometry (at fixed polar angle) between the black hole and cosmological horizons of the four-dimensional Kerr-de Sitter spacetime \cite{Booth:1998gf,Anninos:2009yc,Anninos:2010gh} in the limit that the two horizons coincide, known as the rotating Nariai geometry. This is the de Sitter analogue of the NHEK geometry being geometrically equivalent (at fixed polar angle) to warped anti-de Sitter space.} We should note in passing that additional interesting asymptotically dS$_3$ black hole solutions have been uncovered in `new massive gravity' \cite{Bergshoeff:2009aq,Oliva:2009ip}.


By considering the thermodynamics of the warped dS$_3$ black hole geometries, we find that the entropy of the cosmological horizon can be written in a suggestive form
\be
S_{c} = \frac{\pi^2 \ell}{3} \left( c_R T_R + c_L T_L \right)~.
\ee
The above form resembles a Cardy formula if $T_L$ and $T_R$ are regarded as the left and right moving temperatures of a putative thermal state and $c_L$ and $c_R$ are the right and left-moving central charges of a two-dimensional conformal field theory. Geometrically, $T_L$ and $T_R$ are the coefficients parameterizing the Killing direction which is identified to obtain the black hole solutions. In the limit that $T_R$ vanishes the two horizons coincide, thus defining a Nariai limit \cite{Nariai}. It is notable that the cosmological horizon entropy for warped dS$_3$ black holes is \emph{unbounded}. That is, we can vary the parameters of the solution and find an arbitrarily large entropy. In contrast, the cosmological horizon entropy for black hole configurations in higher dimensional de Sitter space is bounded and attains its maximal value for pure de Sitter space.

In order to find further evidence that there is a dual conformal field theory, we are led to study the asymptotic symmetries of our background. It is found that warped dS$_3$ has an asymptotic symmetry group given by the product of a Virasoro algebra and a current algebra. The central charge and $U(1)$ level are given respectively by
\be
c_R  = \frac{\left(15  \mu ^2 \ell^2 - 81 \right)}{\left(27 - \mu^2 \ell^2 \right)\mu G }~, \quad k =   \frac{ {N'}^2 \, \mu \ell^2}{G\,  (\mu^2 \ell^2-27)}~,
\ee
where $N'$ is a constant (see Sect. 4).
The value obtained for $c_R$ precisely matches the one in the Cardy formula, however we have not recovered the left-moving central charge. Instead we have recovered a $U(1)$ current algebra which via the Sugawara construction contains another copy of the Virasoro algebra \cite{Blagojevic:2009ek}. In this respect, the situation is analogous to that for warped AdS$_3$ mentioned earlier  \cite{Anninos:2008fx,Compere:2008cv, Compere:2009zj}.

We have computed the conserved 
charges 
at $\mathcal{I}^+$ and found a mass gap between the right moving energy of the massless spinless black hole solution and the global background. This differs from the warped anti-de Sitter case, where the background is {\itshape not} recovered for particular values of the mass and angular momentum. If we choose the massless spinless black hole to have vanishing $L_0$ charge, then the mass gap is given by
\be
\Delta E_R = -\frac{c_R}{24\ell}~,
\ee
which is, curiously, in precise agreement with the right moving mass gap between the RR and NS-NS vacua in a conformal field theory \cite{Maldacena:1998bw}.

Our observations are intriguing. We have found clear indications of a holographic duality between warped dS$_3$ and a conformal field theory. But (fortunately) there remain many open questions regarding the nature of the dual field theory. Firstly it is unclear whether the theory is unitary. The conformal weights of the putative operators dual to massive scalar fields about this background are generally complex \cite{Anninos:2009jt}. This may indicate that the theory is non-unitary, or that the scalars are dual to operators transforming under non-highest weight representations such as the principal series representations \cite{Guijosa:2003ze}. Secondly, a thermal density matrix in the bulk is generally obtained by tracing out the unobservable set of modes. In anti-de Sitter space, this process has a natural interpretation at the boundary \cite{Maldacena:2001kr}. It is not clear to us what the boundary theory interpretation of the tracing out of modes is for warped dS$_3$. Thus, it is a pleasant and rather surprising observation that the Cardy formula works.

Finally, one would like to find embeddings of warped dS$_3$ space in consistent theories of quantum gravity, such as string theory, which is clearly a challenging task \cite{Dong:2010pm}.

\section{Framework and Geometry}

We begin our story with the action of topologically massive gravity. It is given by the sum of the Einstein-Hilbert action endowed with a positive cosmological constant and a gravitational Chern-Simons piece,
\be
I_{TMG} = \frac{1}{16\pi G}\int d^3x \sqrt{-g}\left[  R - \frac{2}{\ell^2} + \frac{1}{2\mu}\varepsilon^{\lambda \mu \nu} \Gamma^\rho_{\lambda \sigma} \left( \partial_\mu \Gamma^\sigma_{\rho\nu} + \frac{2}{3} \Gamma^\sigma_{\mu\tau} \Gamma^{\tau}_{\nu \rho} \right) \right]~.
\ee
In what follows we will normalize the positive Newton constant to $G = +1$.

The Chern-Simons term has the effect of adding a single propagating degree of freedom to the theory and causes the usual cosmological Einstein equations to include a term proportional to the Cotton tensor,
\be \label{eom}
R_{\mu\nu} - \frac{1}{2}R g_{\mu\nu} + \frac{1}{\ell^2} g_{\mu\nu} + \frac{1}{\mu}C_{\mu\nu} \equiv \mathcal{G}_{\mu\nu} + \frac{1}{\mu}C_{\mu\nu} = 0~,
\ee
where
\be
C_{\mu \nu} = {{\varepsilon_{\mu}}^{\alpha \beta}}\nabla_{\alpha}(R_{\beta\nu} - \frac{1}{4}g_{\beta\nu}R)~.
\ee
The traceless symmetric Cotton tensor $C_{\mu\nu}$ is the three-dimensional analogue of the Weyl tensor in the sense that it is unaffected by conformal transformations of the metric. Its contribution to the equations of motion leads to a new rich class of non-Einstein solutions. We will now describe the geometry we will focus on.

\subsection{Warped de Sitter Space}

The geometry we will study is given by a real line fibration over two-dimensional de Sitter space. In \emph{global coordinates} the metric is given by
\be
ds^2 = \frac{\ell^2}{3-\nu^2} \left[ -\frac{dt^2}{1+t^2} + (1+t^2)d\phi^2 + \frac{4\nu^2}{(3-\nu^2)}\left( du {+}  t d\phi \right)^2 \right]~,
\ee
where $(t,u) \in \mathbb{R}^2$, $\phi \sim \phi + 2\pi$ and we have defined
\be
\nu \equiv \mu \ell/3
\ee
for convenience. The coordinate $\phi$ is identified in order for our coordinate system to be a single cover of dS$_2$. We would like to emphasize that the above solution displays similar features to de Sitter space only when $\nu^2 < 3$, since when $\nu^2 > 3$ it becomes a real line fibration of AdS$_2$, known as warped AdS$_3$. In what follows, we will focus strictly on the $\nu^2 < 3$ case and we will refer to this geometry as warped de Sitter space or warped dS$_3$ for short. 

The isometries of warped dS$_3$ are given by an $SL(2,\mathbb{R}) \times U(1)$ where the $U(1)$ is non-compact. They simply reflect the isometries of dS$_2$ and translations along the fiber direction $u$. Explicitly, in the above coordinates we find
\begin{eqnarray}
\tilde{J}_1 &=& 2\cos \phi  \sqrt{1 + t^2}\partial_t - 2 \sin \phi \frac{t}{ \sqrt{1 + t^2}} \partial_\phi - \frac{2 \sin \phi}{ \sqrt{1 + t^2}} \partial_u~, \\
\tilde{J}_2 &=& 2\sin \phi   \sqrt{1 + t^2} \partial_t + 2 \cos \phi  \frac{t}{ \sqrt{1 + t^2}} \partial_\phi + \frac{2 \cos \phi}{\sqrt{1 + t^2}} \partial_u~, \\
\tilde{J}_0 &=& 2\partial_\phi; \quad J_2 = 2\partial_u~,
\end{eqnarray}
where the $\tilde{J}_i$ generate the $SL(2,\mathbb{R})$ and $J_2$ generates translations along $u$.

\subsubsection{The Static Patch}

We can also write the metric in the \emph{static patch} which, unlike the above global patch, is the patch accessible to a given observer,
\be
ds^2 = \frac{\ell^2}{3-\nu^2}\left[ -dt^2(1-r^2) + \frac{dr^2}{(1-r^2)} + \frac{4\nu^2}{(3-\nu^2)}\left( du + r dt \right)^2 \right]~.\label{wdsmetric}
\ee
Here $(t,u) \in \mathbb{R}^2$ and $r^2 < 1$ for the patch within the cosmological horizon and $r^2 > 1$ for the patch between the horizon and $\mathcal{I}^\pm$. Several interesting features become clear in this patch. Given that the observers are surrounded by a cosmological horizon at $r^2 = 1$, they reside in a thermal bath of Hawking radiation with temperature:
\be
T_{\text{WdS}_3} = \frac{\sqrt{3 - \nu^2}}{2\pi \ell}~.
\ee
Due to the non-compactness of the spacelike fiber coordinate, the area of the cosmological horizon is in fact infinite and in this sense our spacetime differs from the usual de Sitter case where the cosmological horizon has finite entropy.

Of all possible (higher dimensional) black hole solutions which are asymptotically de Sitter space, it is known that pure de Sitter space carries the most entropy and thus one does not find a spectrum of black holes with arbitrarily large entropy. In the case of warped dS$_3$, given that the static patch has infinite entropy, one may be tempted to look for a spectrum of thermal solutions with arbitrarily large entropy as we will now proceed to do.
\begin{figure}[h]\label{phase}
\centerline{
\includegraphics[width=10cm]{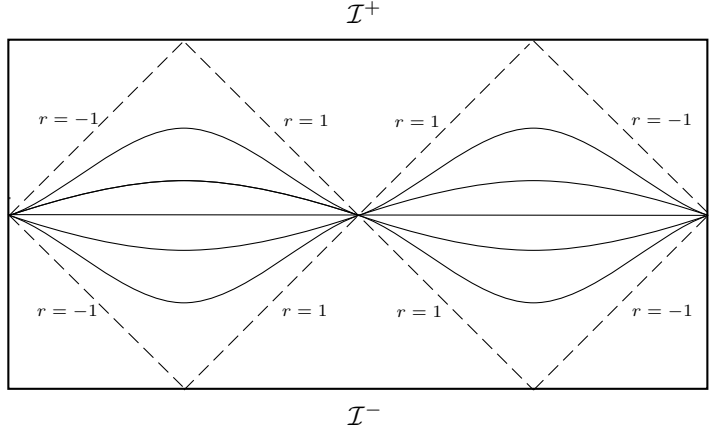}}
\caption{Penrose diagram of dS$_2$. There are cosmological horizons at $r^2 = 1$. Future and past infinity lie at $\mathcal{I}^\pm$ and the left and right edges are identified. The static patch covers anyone of the diamonds, whereas the global patch covers the whole diagram.}
\end{figure}

\section{Asymptotically Warped dS$_3$ Geometries}

We will now discuss a family of asymptotically warped dS$_3$ solutions. These are obtained by considering discrete global identifications of warped dS$_3$.

\subsection{Black Holes}

The black hole solutions are a two parameter family of solutions which are given by global identifications of warped dS$_3$ along a Killing direction which we will express in the following way \cite{Anninos:2008fx}:
\be\label{kill}
\partial_{\tilde \theta} = \pi\ell \left( T_L J_2 + T_R \tilde{J}_2 \right)~,
\ee
where $T_L$ and $T_R$ will eventually find an interpretation as the left and right moving temperatures of the thermal state in the putative dual conformal field theory.

We can build a coordinate system beginning with the global metric and defining a new radial coordinate $\tilde{r} = \alpha \cos \phi \cosh\tau  \equiv \alpha \cos\phi \sqrt{1+t^2}$ ($\alpha$ is a constant), a time coordinate $\partial_{\tilde{t}} = J_2$ and an angular coordinate $\partial_{\tilde{\theta}} = \pi\ell \left( T_L J_2 + T_R \tilde{J}_2 \right)$. Upon computing the norm squared of $\partial_{\tilde{\theta}}$ we find that it remains everywhere positive as long as:
\be\label{bound1}
T_L^2 > \frac{3(1+\nu^2)}{4\nu^2} T_R^2~.
\ee
In the case that it doesn't remain positive everywhere it was shown in \cite{Anninos:2009jt} that there exist naked CTCs rendering the solution unphysical. Furthermore, we can compute the zeroes of $N^2 |\partial_{\tilde{\theta}}|^2$ where:
\be
N^2 = \frac{J_2 \cdot \partial_{\tilde{\theta}}}{|\partial_{\tilde{\theta}}|^2} - |J_2|^2~
\ee
is the lapse function in the ADM decomposition. The zeroes turn out to be located at $r = \pm 1/\alpha \equiv \pm r_h$ and correspond to the horizons.

The quotient constructed above was first obtained as a solution in \cite{Nutku:1993eb,Gurses} and subsequently discussed extensively in \cite{Bouchareb:2007yx}. It is described by the following metric:
\begin{multline}
\frac{ds^2}{\ell^2} =  \frac{2\nu}{(3-\nu^2)^2}
\left( 4\nu  \tilde{r} + \frac{ 3(\nu^2+1)\omega}{\nu {{}} } \right)
d\tilde t d\tilde \theta + \frac{d\tilde{r}^2}{(3 - \nu^2)(r_h - \tilde{r})(\tilde{r}+r_h)} \\ + \frac{4\nu^2}{(3-\nu^2)^2}d\tilde t^2 + \frac{ 3(\nu^2 + 1)}{ {{(3-\nu^2)^2}}} \left( \tilde{r}^2 + 2 \tilde{r} \omega + \frac{(\nu^2-3)r_h^2}{3(\nu^2+1)}  + \frac{3(\nu^2 + 1)\omega^2}{4\nu^2} \right)d\tilde \theta^2~, \label{wdsbc}
\end{multline}
where $\tilde \theta \sim \tilde \theta + 2\pi$. Note that the $\tilde{\theta}$ and $\tilde{t}$ have been rescaled with respect to (5.7) of \cite{Anninos:2009jt} in order to have the same asymptotic form as the vacuum. The black hole and cosmological horizons lie at $-r_h$ and $r_h$ respectively and one can show they have different Hawking temperatures.

In order for the above solution to be free of naked closed timelike curves the parameters must be restricted to obey
\be\label{bound2}
\omega^2 > \frac{4 \nu^2 r_h^2}{3(\nu^2 + 1)}~.
\ee
We can obtain an expression for $T_L$ and $T_R$ in terms of the black hole parameters by considering our above construction mapping the black holes to the vacuum solution. By comparing the terms in $|\partial_{\tilde{\theta}}|^2$ with those in $g_{\tilde{\theta}\tilde{\theta}}$ for instance, we find:
\be
T_L = \frac{3(1 + \nu^2)\omega}{8\pi\ell \nu^2}~, \quad T_R =  \frac{r_h}{2\pi\ell}~.
\ee
Clearly then, (\ref{bound1}) translates into $(\ref{bound2})$ as expected.

The above solution (\ref{wdsbc}) asymptotes to global warped dS$_3$ in the limit $\tilde{r} \to \pm \infty$ and preserves a $U(1)\times U(1)$ isometry. Furthermore, one can recover the global warped dS$_3$ metric upon letting $\omega = 0$ and $r_h = i$. Thus, there is a gap in the parameter space between $r_h^2 > 0$ and $r_h^2 = -1$, reminiscent of the relation between global AdS$_3$ and the BTZ black hole spectrum \cite{Maldacena:1998bw}.


\subsection{Thermodynamics}

We would now like to discuss the conserved charges of the black hole solutions.  In the case at hand, as for regular de Sitter space, there is no natural spacelike infinity. Instead what one finds is that the natural infinities reside at future or past infinity $\mathcal{I}^\pm$ which for each fixed $\tilde{t}$ are given by a spacelike circle. 

Using the 
Barnich-Brandt-Comp\`ere formalism (see Appendix and \cite{Bouchareb:2007yx,Abbott:1982jh,Deser:2002jk,Deser:2002rt})
, we find that the conserved charges at $\mathcal{I}^+$ corresponding to the $\partial_{\tilde{t}}$ and $\partial_{\tilde{\theta}}$ Killing vectors are given by
\be
Q_{\partial_{\tilde{t}}} = \frac{(\nu^2 + 1)}{4 \nu \left( 3 - \nu^2 \right)} \omega~, \quad Q_{\partial_{\tilde \theta}} = \frac{{3} (1+\nu^2)^2}{32 \nu^3 (3 - \nu^2)}\omega^2 - \frac{(5\nu^2 - 3)}{24 \nu (3-\nu^2)}r_h^2~.
\ee

The thermodynamic entropy of the cosmological horizon receives a contribution from the gravitational Chern-Simons term in the action \cite{Tachikawa:2006sz,Kraus:2005zm,Park:1998qk}. We find it convenient to write it in the following form
\be
S_{c} = \frac{\pi^2 \ell}{3} \left( T_L c_L + T_R c_R \right)~,\label{entropy}
\ee
where $T_L$ and $T_R$ are the coefficients appearing in the identified Killing direction (\ref{kill}) and $c_L$ and $c_R$ are two numbers that are independent of the black hole parameters and given by
\be
c_L = \frac{4\nu \ell}{(3-\nu^2)}~, \quad c_R = \frac{(5\nu^2 - 3)\ell}{\nu(3 - \nu^2)}~.
\ee
In section \ref{ASG} we will derive $c_R$ from the asymptotic symmetry group of warped dS$_3$ and discuss the possible meaning of $c_L$.

From the left and right moving temperatures it follows that we can also write expressions for the left and right moving energies
\be
E_{L} = \frac{\pi^2\ell}{6} c_L T_L^2~, \quad E_{R} = \frac{\pi^2\ell}{6} c_R T_R^2~,
\ee
as is usual in a two-dimensional conformal field theory with left and right moving central charges $c_L$ and $c_R$. We note that the angular momentum is indeed given by the difference between the right and left-moving temperature,
\be
Q_{\partial_{\tilde{\theta}}} = E_L - E_R
\ee
in analogy with the BTZ black hole case.

As a final consistency check, one can show that the first law of thermodynamics is satisfied for the cosmological horizon,
\be
\delta Q_{\partial_{\tilde{t}}} = {T}_H \delta S_{c} + {\Omega}_c \delta Q_{\partial_{\tilde \theta}}~,
\ee
where
\be
T_H = \frac{1}{\pi \ell}\frac{2 r_h \nu^2}{\left( 4 r_h \nu^2 + 3 (1 + \nu^2) \omega \right)}~, \quad \Omega_c = \frac{4 \nu^2}{\ell(4  r_h \nu^2 + 3 \omega + 3  \nu^2 \omega)}
\ee
are the Hawking temperature and angular velocity of the cosmological horizon at $r_h$. Having said this, we have completed our semi-classical discussion of the black hole solutions.

\subsection{Nariai Limit}

One can find the Nariai limit of the black holes by focusing on the region of the geometry between the cosmological and black hole horizons in the limit where they coincide. In this limit, $r_h \to 0$ and thus the right moving temperature vanishes. We can define an appropriate set of near horizon coordinates
\be
\tilde{r} \to (r_h -\epsilon r), \quad t \to \frac{t}{\epsilon}~, \quad \tilde{\theta} \to \tilde{\theta} - \frac{\Omega_c t}{\epsilon}~, \quad \Omega_c \equiv \frac{g_{t\tilde \theta}}{g_{\tilde \theta\tilde \theta}} \mid_{\tilde{r}=r_h}~,
\ee
and take the near horizon limit $\epsilon \to 0$ with $r_h = \epsilon \tilde{r}_h$. We are thus lead to the following near horizon geometry
\be \label{NHG}
ds^2 = \frac{\ell^2}{3-\nu^2} \left[ -(\tilde{r}^2_h - r^2){dt^2} + \frac{dr^2}{(\tilde{r}^2_h - r^2)} + \frac{4\nu^2}{(3-\nu^2)}\left( d\tilde \theta + r dt \right)^2 \right]~,
\ee
where $\tilde{\theta} \sim \tilde{\theta} + 2\pi \beta$ and $\beta = 2\pi T_L$. The $U(1)\times U(1)$ isometry group is enhanced to an $SL(2,\mathbb{R})\times U(1)$ in the Nariai limit. Notice that when $\tilde{r}_h \neq 0$, both horizons are preserved in the near horizon limit.

When $\tilde{r}_h=0$, the above geometry is simply the vacuum solution with the fiber coordinate identified, i.e. self-dual warped dS$_3$ in planar coordinates. It would be very interesting to understand the asymptotic symmetry group of this near horizon geometry and in particular obtain a left moving central charge $c_L$ from such considerations. Finally, we should emphasize that the metric (\ref{NHG}) is obtained (at fixed polar angle) from the Nariai limit of a rotating black hole in a four-dimensional de Sitter universe \cite{Booth:1998gf,Anninos:2009yc,Anninos:2010gh}.

\section{Asymptotic Structure}\label{ASG}

As discussed earlier, the form of the warped dS$_3$ black hole metrics is closely related to that of the spacelike warped AdS$_3$ black holes discussed in \cite{Anninos:2008fx}. In particular, their asymptotic structure is related by an analytic continuation. One can see this by examining the metric for warped AdS$_3$ as a solution to TMG with negative cosmological constant in the global patch
\be
ds^2 = \frac{\ell^2}{3+\nu^2} \left[ -(1+r^2)d\tau^2 + \frac{dr^2}{1+r^2} + \frac{4\nu^2}{(3+\nu^2)}\left( du +  r d\tau \right)^2 \right]~. \label{wads1}
\ee
Clearly, we can obtain warped dS$_3$ space in global coordinates from (\ref{wads1}) by analytically continuing $\nu \to i\nu$ and $\ell \to i\ell$ and relabeling $r$ with $t$ and $\tau$ with $\phi$.

\subsection{Boundary Conditions}

Boundary conditions and the asymptotic symmetry algebra for warped AdS$_3$ geometries
have been discussed in \cite{Compere:2007in,Compere:2008cv,Compere:2009zj,Blagojevic:2009ek}.

In view of their similarities with the geometries under study here, one can devise the corresponding boundary conditions for warped dS$_3$ at $\tilde r \rightarrow \infty$, along the lines of \cite{Compere:2009zj}.
Thus, we only allow as part of our phase space variations of the metric that behave as follows at future infinity:
\bea \label{BC}
\delta g_{\tilde t \tilde t} &\sim& 1/\tilde{r}~, \quad \delta g_{\tilde t \tilde r} \sim 1/\tilde{r}^2~, \quad \delta g_{\tilde t \tilde{\theta} } \sim 1~,\\
\delta g_{\tilde r \tilde r} &\sim& 1/\tilde{r}^3~, \quad \delta g_{\tilde r \tilde{\theta} } \sim 1/\tilde{r}~, \quad \delta g_{\tilde{\theta} \tilde{\theta} } \sim \tilde{r}~. \notag
\eea
The above boundary conditions define asymptotically warped dS$_3$ spaces and allow for all the solutions discussed in the present work.



\subsection{Asymptotic Symmetry Group}

The asymptotic symmetry group of warped dS$_3$ is given by the set of diffeomorphisms preserving the above boundary conditions and yielding well-defined charges, quotiented by those trivial diffeomorphisms with vanishing charges. We find a semi-direct product of a Witt algebra with a $\widehat{u(1)}$ current algebra, satisfying
\begin{eqnarray}
i[l_m,l_n] &=& (m-n)l_{m+n}~, \\
i[l_m,t_n]&=& -n t_{m+n}~,\qquad [t_m,t_n] = 0~,\label{alg}
\end{eqnarray}
generated by
\begin{eqnarray}
l_m  &=& N e^{i m \tilde \theta} \partial_{\tilde{t}} - i \tilde{r} m   e^{i m \tilde \theta} \partial_{\tilde{r}} + e^{i m \tilde \theta}  \partial_{\tilde \theta}~, \\
t_m &=& N'  e^{i m \tilde \theta}  \partial_{\tilde{t}}~,
\end{eqnarray}
where $N$ and $N'$ are normalization constants which are unspecified classically.
The subset $(l_{-1},l_0,l_1,t_0)$ forms a $sl(2,\mathbb R) \oplus \mathbb R$ subalgebra which consists of the exact background Killing vectors.

\subsection{Generators}

Using covariant phase space techniques \cite{Barnich:2001jy,Barnich:2007bf} we can compute the charges associated with the asymptotic symmetries. These charges generate the asymptotic symmetries. Given a background metric configuration $\bar{g}_{\mu\nu}$,
the infinitesimal charge difference between $\bar{g}_{\mu\nu}$ and $\bar{g}_{\mu\nu} + \delta g_{\mu\nu}$ is given by:
\be
\delta \mathcal{Q}_\xi \left[ \bar{g},\delta g \right] = \int_{\partial \mathcal{M}} \sqrt{-\bar{g}} k_\xi^{\mu\nu} \left[\delta g, \bar{g} \right] \epsilon_{\mu \nu \rho} dx^\rho~.
\ee
The one-form $k_\xi^{\mu\nu} \left[\delta g, \bar{g} \right] \epsilon_{\mu \nu \rho} dx^\rho$ is
constructed out of the equations of motion of the theory and is given explicitly in \cite{Compere:2008cv,Compere:2009zj}.\footnote{Explicit formulas are also provided in the appendix.} The integral is over the boundary circle $\partial \mathcal{M}$.
Computing the Poisson bracket algebra for our charges we find:
\begin{eqnarray}
i \left\{ L_m, L_n \right\} &=& \left( m - n \right) L_{m+n} + \frac{c}{12} m \left( m^2 - 1 \right)\delta_{m+n}~,\\
i \left\{ L_m, T_n \right\} &=& - n T_{m+n} + \frac{N N' \nu}{3(\nu^2-3)}
~,\\
i \left\{ T_m, T_n \right\} &=&  \frac{{N'}^2 \, \ell \nu }{3 G\,  (\nu^2-3)}
m \delta_{m+n}~,
\end{eqnarray}
where we have defined $L_n \equiv \mathcal{Q}_{l_n}$ and $T_n \equiv \mathcal{Q}_{t_n}$. Thus, the asymptotic symmetry group acquires a central extension at the level of the charges. The central charge is given by:
\be
\mathcal{K}_{\xi,\xi'} = \int_{\mathcal{\partial \mathcal{M}}} k_\xi \left[ \mathcal{L}_{\xi'} \bar{g}, \bar{g} \right]~.
\ee
We obtain the following values for the Virasoro central charge $c$ and the $\widehat{u(1)}$ level $k$:
\begin{eqnarray}
c = \frac{(5 {\nu}^2 - 3)\ell}{ {\nu}(3-{\nu}^2 )G} \equiv c_R~, \quad k =   \frac{{N'}^2 \, \ell \nu }{3 G\,  (\nu^2-3)}~.
\end{eqnarray}
It is worth noting at this point that there exists a left-moving Virasoro given by applying the Sugawara construction to the affine algebra \cite{Blagojevic:2009ek}. Though appealing, this procedure proves insufficient to fix the precise value of $c_L$.


\section{Conjecture}

The above discussion leads us to the natural conjecture that quantum gravity in asymptotically warped dS$_3$ is holographically dual to a two-dimensional conformal field theory living at $\mathcal{I}^+$ with central charges given by
\be
c_L = \frac{4\nu\ell}{(3-\nu^2)G}~, \quad c_R = \frac{(5\nu^2 - 3)\ell}{\nu(3 - \nu^2)G}~.
\ee
Our evidence rests in the analysis of the asymptotic symmetries of the space time which are comprised of a Virasoro algebra in addition to an affine algebra with non-vanishing level. 
As we have already mentioned, it may be possible to obtain the second Virasoro from a Sugawara construction, as in \cite{Blagojevic:2009ek}. Another convincing piece of evidence would be to recover $c_L$ for the self-dual geometry (\ref{NHG}) along the lines of \cite{Chen:2010qm,Anninos:2008qb,Balasubramanian:2009bg}.

Assuming the usual growth of states for a two-dimensional CFT \cite{Cardy:1986gw}, one can use the Cardy formula to predict the entropy of the thermal states leading to a natural interpretation of the black hole entropy formula in (\ref{entropy}). Thus, asymptotically warped dS$_3$ black hole spacetimes are the natural bulk manifestations of the thermal states in the CFT. We should note that the entropy predicted by the Cardy formula gives only the entropy of the cosmological horizon and not the black hole horizon at $-r_h$.\footnote{One could study the asymptotic structure as $\tilde{r} \to -\infty$, i.e. the infinite region behind the inner horizon at $-r_h$. It is conceivable that there exists a similar structure accounting for the remaining entropy. We should note, however, that the entropy of the black hole horizon takes the form $S_{BH} = \frac{\pi^2 \ell}{3}\left( c_L T_L - c_R T_R \right)$ and thus the product $S_{BH} S{_c} = \pi^2 \left(c_L E_L - c_R E_R\right)/3$ is an integer. In line with the discussion of \cite{Cvetic:2010mn}, this might be interpreted as a level matching condition.}

Switching off the right moving temperature $T_R$, we are left with the Nariai geometry. Consequently, deviations away from the Nariai limit are obtained by turning on $T_R$. Therefore, the operational meaning of varying the size of the cosmological horizon corresponds to tuning the right moving temperature in the putative holographic dual.\footnote{It would be interesting to understand whether this observation carries forward to the rotating Nariai/CFT correspondence studied in \cite{Anninos:2009yc,Anninos:2010gh}. Our considerations might suggest that turning on a right moving temperature corresponds to leaving the rotating Nariai limit in which the cosmological and black hole horizons coincide.}

Furthermore, there is a gap in the right moving energy between the $r_h = \omega = 0$ black hole and the ground state given by
\be
\Delta E_R = -\frac{c_R}{24\ell}~.
\ee
This resembles the mass gap found between the massless spinless BTZ black hole and global AdS$_3$ \cite{Maldacena:1998bw,Banados:1992gq,Banados:1992wn}.

Finally, the right moving central charge vanishes at the value
\be
\nu^2 = 3/5~.
\ee
It is very tempting to consider that TMG in a warped dS$_3$ background at $\nu^2 = 3/5$ is a chiral theory given that the right moving energy $E_R$ vanishes in this limit and consequently, the entropy only depends on $E_L$. Also, the black hole and cosmological entropy are equal for this value of $\nu$. We leave these observations for future work. To provide evidence for such a proposal, that the linearized spectrum and stability properties of the warped dS$_3$ vacua must be carefully analyzed.

\section{Instantons}

In this brief section, we mention that we can construct an instanton that interpolates between the no boundary condition and the self-dual warped dS$_3$ geometry. This instanton may mediate the production of black holes in the limit where both the black hole and cosmological horizons coincide.

It is convenient to write the global metric in the following coordinate system
\be
ds^2 = \frac{\ell^2}{3-\nu^2} \left[ -d\tau^2 + \cosh^2\tau d\phi^2 + \frac{4\nu^2}{(3-\nu^2)}(du + \sinh\tau d\phi)^2\right]
\ee
by performing the coordinate transformation $t = \sinh\tau$. We can construct the instanton by analytically continuing $\tau$ to $i \theta$ and periodically identifying $u$. The resulting compact Euclidean geometry is given by
\be
ds^2 = \frac{\ell^2}{3-\nu^2} \left[ d\theta^2 + \cos^2\theta d\phi^2 + \frac{4\nu^2}{(3-\nu^2)}(du + i \sin\theta d\phi)^2\right]~,
\ee
where $\theta \in [-\pi/2,\pi/2]$, $u \sim u + 2\pi\beta$ and $\phi \sim \phi + 2\pi$. Notice that the above metric is complex. We can make it real by further analytically continuing $u$ to $i u$ and $\nu$ to $i \nu$ after which the resultant geometry is found to be a quotient of the squashed three-sphere. However, as discussed in \cite{Booth:1998gf}, we will retain the complex solution since that is the one that matches onto the Lorentzian solution.

Thus, one can consider the Euclidean instanton with no boundary condition at $\theta = -\pi/2$ glued to the Lorentzian solution at $\theta = 0$ where the two geometries match as a process where self-dual warped dS$_3$ is produced starting with nothing. The process we have described resembles the cosmological pair production of four-dimensional Schwarzschild-de Sitter black holes in the Nariai limit (where the black hole and cosmological horizons coincide) \cite{Bousso:1996au}. One difference however is that whereas one can find an instanton that can be matched to Lorentzian de Sitter space, we do not know of a compact instanton that can be matched onto the full global warped dS$_3$ and it is not clear that such an instanton even exists.

It is tempting to use the Euclidean quantum gravity prescription for the Hartle-Hawking wavefunction to compute the relative probability for nucleation of de Sitter versus warped dS$_3$ universes \cite{Bousso:1996au} by computing on-shell actions. We reserve such temptations for the future due to the subtle intricacies of analytically continuing Chern-Simons theories \cite{Witten:2010cx}.

\section*{Acknowledgements}

It has been a great pleasure discussing this work with Tarek Anous, Alejandra Castro, Geoffrey Comp\`ere, Alex  Maloney and Andy Strominger. D.A. was supported in part by DOE grant DE-FG02-91ER40654. The works of S.dB. and S.D. are funded by the European Commission through the grants PIOF-GA-2008-220338 and PIOF-GA-2008-219950. S.dB. and S.D. thank Harvard University for hospitality while this work was initiated.

\appendix

\section{BBC formalism for conserved charges}

We sketch the general ideas and methods for constructing (asymptotically) conserved charges in the formalism of Barnich, Brandt and Comp\`ere. We refer the reader to references \cite{Barnich:2001jy, Barnich:2007bf, Compere:2006my, Compere:2007az, Barnich:2003xg} for further details and references.

There are well-known puzzles in gauge theories when it comes to defining conserved quantities associated with symmetries. Take as an example general relativity admitting diffeomorphism invariance: $x^\mu \rightarrow x^\mu + \xi^\mu$, $\; g_{\mu \nu} \rightarrow g_{\mu \nu} + \cL_{\xi}g_{\mu \nu}$, for an arbitrary vector field $\xi$. It is clear that not every $\xi$ will yield a conserved quantity. The associated canonical Noether current can be written as:
\begin{equation}
J^\mu_\xi = 2 \sqrt{-g} G^{\mu \nu} \xi_\nu + \p_\nu k^{[\mu \nu]}~,
\end{equation}
for some skew-symmetric $k^{\mu \nu}$. However, the Noether current is only defined up to terms vanishing on-shell and up to a closed $(n-2)$ form (see (3) of \cite{Compere:2007az}), and could therefore have been chosen to be zero. The idea is that conservation laws in gauge theories are {\itshape lower degree conservation laws}, i.e. based on the conservation of an $(n-2)$-form $k = k_{\mu \nu} (d^{n-2} x)_{\mu \nu}$ with $dk = \p_\nu k_{\mu \nu} (d^{n-1} x)_\mu \approx 0$ instead of that of an $(n-1)$-form $J = J^\mu  (d^{n-1} x)_\mu$. The charge associated with $\xi$ could then be defined as $Q_\xi = \oint_{S^\infty} k_\xi^{\mu \nu} (d^{n-2}x)_{\mu \nu}$, but Noether's theorem does not provide an unambiguous method to determine $k_{\mu \nu}$, which is totally arbitrary at this point.

A convenient framework to deal with $p-$form conservation laws relevant to gauge theories is field theoretical local cohomology. An important ingredient is the so-called characteristic cohomology $H^{n-p}_{char} (d)$ of on-shell closed $(n-p)$-forms ($d\omega \approx 0$) modulo on-shell exact forms ($\omega \approx d \omega'$). The fact that $k$ is closed implies that the corresponding charge does not depend on time and that the integration domain can be freely deformed in vacuum regions. For $p=1$, one gets the cohomology of non-trivial conserved currents that can be shown to be equal to the cohomology of global symmetries of the theory. This is just Noether's theorem expressed in a fancy way.

For general relativity, this cohomology is trivial as a consequence of the non-existence of non-trivial global symmetries. The definition of charges in gauge theories relies on the {\itshape generalized Noether theorem}, stating there exists a one-to-one correspondence between equivalence classes of $(n-2)$-forms that are conserved on-shell and non-trivial reducibility parameters of the theory.\footnote{These are parameters of a gauge transformation vanishing on-shell such that the parameter itself is non zero on-shell. For example, in electromagnetism the trivial gauge transformations $\delta A_\mu = \p_\mu c \approx 0$ correspond to constants, so there exists a single reducibility parameter up to a multiplicative constant that can be absorbed, and the corresponding conserved $(n-2)$-form gives the electric charge.}

For Einstein gravity (or for generally covariant theories), no such reducibility parameters exist: no vector is a Killing vector of all solutions to Einstein's equations. As a consequence, there is no general formula for a non-trivial conserved $(n-2)$-form locally constructed from the metric. This does not prevent the definition of conserved quantities for restricted classes of spacetimes such as those admitting Killing vectors or a common asymptotic structure. The key there is to use the linearized theory around some background $\bar g$, for which the gauge transformations read $\delta_\xi h = \cL_\xi g$. This yields reducibility parameters when the reference field admits symmetries, either as the first order approximation when performing infinitesimal field variations, or as an approximation to the full theory at the infinite distance boundary.

Charges of the full theory can then be obtained by integrating the infinitesimal charge differences (see e.g. \cite{Compere:2007az} p14). In the former case, one applies the linearized theory around a family of solutions $g_{\mu \nu}$ having an exact Killing vector $\xi$ to compute the charge difference between $g_{\mu \nu}$ and  $g_{\mu \nu} + \delta g_{\mu \nu}$. The total charge associated with $\xi$ is then
\begin{equation}
\cQ_\xi[g ;\bar g] = \int_{\bar g}^g \int_S \sqrt{-g}\, k^{\mu\nu}_\xi[\delta g ; g]    \eps_{\mu\nu\rho}\,dx^\rho
\end{equation}
where $\bar g$ is a background solution with charge normalized to zero and the inner integral is performed along a path of solutions. The generalized Noether theorem can further be generalized to account for {\itshape asymptotic symmetries}: there is a one-to-one correspondence between equivalence classes of asymptotic irreducibility parameters and equivalence classes of asymptotically conserved $(n-2)$ forms (see e.g. p8 of \cite{Barnich:2001jy}).

The conserved $(n-2)$-forms are obtained as
\begin{equation}\label{k}
   k^{BB}_\xi[\delta g,g] = I^{n-1}_{\delta g} S_\xi~.
\end{equation}
$S_\xi = 2 \frac{\delta L}{g_{\mu \nu}} \xi_\nu (d^{n-1}x)_\mu$ is the weakly vanishing Noether current for a gravity theory with Lagrangian $L$ and the homotopy operator $I^{n-1}_{\delta g}$ is defined in (A.29) and (A.9) of \cite{Compere:2007az}. They depend solely on the equations of motion of the theory. The surface charges are intimately related to the following invariant presymplectic forms (see Appendix C of \cite{Anninos:2010pm} and \cite{Andrade:2009ae} for the explicit form):
\begin{eqnarray}
W(\delta_1 g,\delta_2 g) = -\frac{1}{2} I^n_{\delta_1 g} \left[(\delta_2 g)^{\mu \nu} \frac{\delta L}{\delta g^{\mu \nu}}\right]~, \quad \Omega^{IW} (\delta_1 g,\delta_2 g) = \delta_1 (I^n_{\delta_2 g} L)~.
\end{eqnarray}
The latter is the one usually used in the context of covariant phase space methods and yields in particular the Iyer-Wald expression for conserved charges in general relativity \cite{Iyer:1994ys}. They are related by
$$
- W = \Omega^{IW} + d_H E
$$
where $E = \frac{1}{2}  I^{n-1}_{\delta_2 g} \Theta(\delta_1 g)$, with $\Theta$ defined through $\delta L = \delta g^{\mu \nu} \frac{\delta L}{\delta g^{\mu \nu}} + \p_\mu \Theta^\mu (\delta g)$ and $\Theta = \theta^\mu (d^{n-1}x)_\mu$. $d_H$ is the horizontal differential defined in (A.3) and (A.1) of \cite{Compere:2007az}. The charge $(n-2)$-form \re{k} can be further rewritten as:
\be
k^{BB}_\xi[\delta g,g] \approx I^{n-1}_{\xi} W[\delta g, \cL_\xi g] + d_H(.)~.
\ee
Using $\{i_{\cL_\xi},d_H\}$=0, one gets $W[\delta g, \cL_\xi g]  = \Omega[\cL_\xi g,\delta g] + d_H E(\delta_g,\cL_\xi g)$.
With  $k^{IW}_\xi[\delta g,g] \approx I^{n-1}_{\xi} W[\cL_\xi g,\delta g]$ and using the basic property (A.37) of \cite{Compere:2007az} for a homotopy operator, one finds
\be
k^{BB}_\xi[\delta g,g]  = k^{IW}_\xi[\delta g,g] + E(\delta_g,\cL_\xi g)~.
\ee
The charges defined from $k^{BB}$ and $k^{IW}$ thus differ by a term vanishing for exact Killing vectors but potentially giving a contribution in the asymptotic case. In most cases this supplementary term has a vanishing contribution (for a notable exception see Sect. 5.4 of \cite{Azeyanagi:2009wf}).

In this paper, we used the Iyer-Wald expression that has been derived in \cite{Bouchareb:2007yx,Compere:2008cv}:\footnote{The Mathematica code implementing that expression can be downloaded from the homepage of G. Comp\`ere: http://www.physics.ucsb.edu/$\sim$gcompere/}
\begin{multline}
(16 \pi G)\,k^{\mu\nu}_{\xi}[\delta g,g] = (16 \pi G)\,k^{\mu\nu}_{Ein,\xi_{tot}}[\delta g,g]
-\frac{1}{\mu\sqrt{-g}}\xi_\lambda \left[ 2 \eps^{\mu\nu\rho}\delta (G^\lambda_{\;\rho}) -  \eps^{\mu\nu\lambda}\delta G \right]  \\
 -\frac{1}{\mu\sqrt{-g}} \eps^{\mu\nu\rho} \left[ \xi_\rho h^{\lambda\sigma}G_{\sigma\lambda} +\frac 1 2 h\left(\xi_\sigma G^\sigma_{\;\rho}+\frac 1 2 \xi_\rho R\right) \right]~,
\label{chargetot}
\end{multline}
where $\delta g_{\mu\nu} = h_{\mu \nu}$, $\xi_{tot}^{\nu} = \xi^\nu + \frac 1 {2\mu \sqrt{-g}} \eps^{\nu \rho\sigma}D_\rho \xi_\sigma$ and $k^{\mu\nu}_{Ein,\xi}[\delta g;g]$ is the Iyer-Wald expression \cite{Iyer:1994ys} for general relativity:
\begin{multline}
\sqrt{-g}\,k^{\mu\nu}_{Ein,\xi}[\delta g,g] = \sqrt{-g}\,\xi^\mu\left(D_\lambda h^{\lambda \nu}-D^\lambda h\right) -\delta\left(\sqrt{-g}D^\mu \xi^\nu\right) - \left(\mu \leftrightarrow \nu\right)~.
\end{multline}
An important property of the (asymptotic) charges so defined is that they form a representation of the (asymptotic) symmetry algebra up to central terms through a covariant Poisson bracket, according to theorem 12 of \cite{Barnich:2007bf}\footnote{The theorem holds modulo a technical assumption on the form of the supplementary term $E$.}:
\begin{equation}
\{Q_\xi[g ;\bar g],Q_{\xi^\prime}[g ;\bar g]\} = Q_{[\xi,\xi^\prime]}[g ;\bar g] + \cK_{\xi,\xi^\prime}[\bar g]~.
\label{eq:repr}
\end{equation}
with
\be
\cK_{\xi,\xi^\prime}[\bar g] \equiv \int_S k^{\mu\nu}_{\xi}[\cL_{\xi^\prime} \bar g;\bar g] (d^{n-2}x)_{\mu \nu}~.
\ee
In particular, when asymptotic symmetries form a Virasoro algebra, the central extension $c$ is given by
\be
\frac{c}{12} m (m^2 + a) \delta_{m+n,0} = i \oint_{S^\infty}  k_{l_m}[\cL_{l_n}\bar g,\bar g]~,
\ee
where $a$ is a constant depending on the normalization of the conventional $L_0$ energy of the background $\bar g$.

\end{document}